\documentclass{elsarticle}
\usepackage[margin=3.4cm]{geometry}
\usepackage{latexsym}
\usepackage{longtable}
\usepackage{float}
\usepackage[latin2]{inputenc}
\usepackage{graphicx}
\usepackage{ae}
\usepackage{aecompl}
\usepackage{setspace}
\usepackage{amsmath}
\usepackage{amssymb}
\usepackage{amsthm}
\usepackage[hidelinks]{hyperref}

\makeatletter
\def\ps@pprintTitle{%
	\let\@oddhead\@empty
	\let\@evenhead\@empty
	\def\@oddfoot{\centerline{\thepage}}%
	\let\@evenfoot\@oddfoot}
\makeatother

\begin{document}

\title{The braingraph.org Database with more than 1000 Robust Human Structural Connectomes in Five Resolutions}

\author[p]{Bálint Varga}
\ead{balorkany@pitgroup.org}
\author[p,u]{Vince Grolmusz\corref{cor1}}
\ead{grolmusz@pitgroup.org}
\cortext[cor1]{Corresponding author}
\address[p]{PIT Bioinformatics Group, Eötvös University, H-1117 Budapest, Hungary}
\address[u]{Uratim Ltd., H-1118 Budapest, Hungary}

\date{}

\begin{abstract}
The human brain is the most complex object of study we encounter today. Mapping the neuronal-level connections between the more than 80 billion neurons in the brain is a hopeless task for science. By the recent advancement of magnetic resonance imaging (MRI), we are able to map the macroscopic connections between about 1000 brain areas.  The MRI data acquisition and the subsequent algorithmic workflow contain several complex steps, where errors can occur. In the present contribution, we describe and publish 1064 human connectomes, computed from the public release of the Human Connectome Project. Each connectome is available in 5 resolutions, with 83, 129, 234, 463, and 1015 anatomically labeled nodes. For error correction, we follow an averaging and extreme value deleting strategy for each edge and for each connectome. The resulting 5320 braingraphs can be downloaded from the \url{https://braingraph.org} site. This dataset makes possible the access to these graphs for scientists unfamiliar with neuroimaging- and connectome-related tools: mathematicians, physicists, and engineers can use their expertize and ideas in the analysis of the connections of the human brain. Brain scientists also have a robust and large, multi-resolution set for connectomical studies. 
\end{abstract}

\maketitle

\section*{Background \& Summary}
Connectomes or braingraphs are compact and focused derivatives of the diffusion magnetic resonance images (MRIs) of the brain: their vertices are labeled by the anatomical areas, and two such vertices are connected by a weighted graph-edge, if a tractography workflow \cite{Besson2014a} finds neural tracks between the areas, corresponding to the vertices. By focusing to the connections between cerebral areas instead of analyzing the whole MR image, we can make use the rich and refined resources of graph theory, born by the famous article of Leonhard Euler on the problem of the Königsberg Bridges \cite{Eulera} in 1741.

Our research group earlier has prepared several undirected and directed braingraph sets \cite{Kerepesi2016b,Szalkai2015a,Szalkai2016,Kerepesi2015b, Szalkai2016d} from the 500 Subjects Data Release \cite{McNab2013} of the Human Connectome Project (HCP). The resulting graphs were made available at the site \url{https://braingraph.org}, and were applied in several structural studies of the human brain \cite{Szalkai2015,Kerepesi2015a, Szalkai2016d, Kerepesi2016, Szalkai2016c, Szalkai2017c, Szalkai2016e, Szalkai2016a,Fellner2017,Fellner2019,Fellner2018,Fellner2019a}. 

In the present contribution, we describe a new braingraph set, computed from the 1200 Subjects Data Release of the Human Connectome Project \cite{McNab2013}. The set contains 1064 connectomes, each in five resolutions, and each edge is weighted by three different weight functions.

\section*{Methods}

The data source of the workflow is the 1200 Subjects Data Release of the Human Connectome Project \cite{McNabb2013}, documented at the site \url{https://www.humanconnectome.org/study/hcp-young-adult/document/1200-subjects-data-release}. For the present study the ``re-preprocessed'' 3T diffusion data (cf. \url{https://www.humanconnectome.org/study/hcp-young-adult/document/1200-subjects-data-release}) was applied. 

The CMTK workflow \cite{Daducci2012} was utilized in the graph computation on the HCP data. For each subject, we have applied the segmentation and the parcellation steps only once, but the probabilistic tractography part of the workflow 10 times. The parcellation scheme was the Lausanne2008 atlas; the labels applied are listed in \url{https://github.com/LTS5/cmp_nipype/blob/master/cmtklib/data/parcellation/lausanne2008/ParcellationLausanne2008.xls}. 

The graph construction was performed in the following steps:
\begin{itemize}
	
\item[1,] For each subject, the MRtrix 0.3 tractography algorithm was run, with probabilistic seeding and probabilistic tractography. The number of streamlines was set to 1 million. For defining the graph edges, let us consider two distinct, anatomically labeled areas of the cortical- or sub-cortical gray areas of the brain, denoted by $A$ and $B$. If the tractography algorithm found at least one streamline between the area  $A$ and $B$, then vertex $a$, representing area $A$ was connected to vertex $b$, representing area $B$, by a graph edge. The three weights of $\{a,b\}$ give the number of streamlines or fibers found between areas $A$ and $B$, the average length of the streamlines, and the mean fractional anisotropy of the streamlines. 

\item[2,] Step 1 was repeated 10 times for each subject. We accepted $\{a,b\}$ to be an edge of the connectome of the subject if it was present in all ten graphs computed in the repetitions. Next, for each edge, we computed the maximum and the minimum number of the fibers, defining that edge, and deleted those two extremal values. Consequently, there remained 8 fiber numbers for each edge. We computed the mean value of those fiber numbers, the mean value of the lengths of the streamlines and the fractional anisotropies for the three weights of the edge.
  
\end{itemize}

In other words, the probabilistic tractography was performed 10 times, the graphs were constructed after each run, (i.e., 10 graphs were constructed for each subject), next the extremal fiber number values were deleted, the remaining 8 values were averaged, and the edges, which were present in all 10 graphs were allowed to be included in the resulting graph. 

Steps 1 and 2 were performed only in the highest (i.e., the finest) resolution with 1015 vertices. For lower resolutions, the graphs were computed from the 1015-vertex graph by contracting vertices, summing the fiber numbers of the multiple edges between the two contracted vertices and contracting the multiple edges.

On the choice of 10 as the repetition number of the probabilistic tractography we refer to the detailed analysis in the ``Technical Validation'' section below.

From the dataset of the HCP website we were able to finish the graph computations for 1064 subjects.

The computation was done on our 24-member Intel i7 cluster (each with 6 physical and 12 virtual CPU cores and 16 GB of RAM) within 3 weeks running time. 

\section*{Data Records}

The data source of this work was published at the Human Connectome Project's website at \url{http://www.humanconnectome.org/} \cite{McNab2013} as the 1200 Subjects Public Release. The parcellation data, containing the anatomically labeled ROIs, is listed in the CMTK nypipe GitHub repository \url{https://github.com/LTS5/cmp_nipype/blob/master/cmtklib/data/parcellation/lausanne2008/ParcellationLausanne2008.xls}. 

The braingraphs, computed by us, can be accessed at the  \url{https://braingraph.org/cms/download-pit-group-connectomes/} site, by selecting one of the download options, denoted by  ``X nodes set, 1064 brains, 1 000 000 streamlines, 10x repeated'', where $X=86, 129, 234, 463, 1015$.

The graphs are given in GraphML format, described in \url{https://cmtk.org} \cite{Daducci2012}. Each file begins with an attribute definition section; then the nodes are described with their coordinates and anatomical labels, corresponding to the parcellation at \url{https://github.com/LTS5/cmp_nipype/blob/master/cmtklib/data/parcellation/lausanne2008/ParcellationLausanne2008.xls}.

Next, the (undirected) edges are listed. The edges carry three weights: 

\begin{itemize}
	\item the number of fibers;
	\item the mean value of the fiber lengths of the fibers, defining the edge;
	\item and the mean fractional anisotropy of the fibers   
\end{itemize}

Note that the edge weights are averages from multiple tractography-runs; therefore, even the fiber number is -- typically -- a non-integer.

\section*{Technical Validation}

Here we describe the workflow, which implied the choice of the 10 repetitions of step 1 in the graph construction above. We note that the present section describes only the process, resulting the specific choice of the repetition number 10, and not the actual graph construction (which was already duly described in the ``Methods'' section).

The implementations of the deterministic tractography algorithms also contain a probabilistic seeding step; i.e., two runs of these tractography computations almost always yield different results. When we use probabilistic tractography \cite{Girard2014,Buchanan2014a}, it is evident that distinct runs yield different results. 

For generating reproducible results in the graph construction with a probabilistic tractography phase, it is a natural idea to repeat the probabilistic tractography algorithm for the very same input several times, and to average the results of the tractography in a careful way. Here we assume that the differences in the number of the discovered fibers between the very same two vertices are distributed randomly; more exactly, we assume that the expectation of these differences is 0. This assumption implies that the repetitions and the averaging will increase the reliability of the tractography results.

For the determination of the number of repetitions $k$, with the trade-off with practical computability and robustness, we have followed the strategy, described as follows. In short, we determined the number of necessary repetitions by comparing deviations for 10 average values, each for $k$ repetitions, for $k=1,2,\ldots,50$. 

More exactly, we have chosen 9 subjects: for each non-zero leading digits of the ID numbers, one was chosen randomly (the choices were: 136631, 200008, 300618, 401422, 500222, 601127,700634, 800941, 901038). For a given subject, and a given positive integer value $k$, we have generated the following ten braingraphs:

$${G_k}_1, {G_k}_2, ... {G_k}_{10},$$

where ${G_k}_i$ was calculated by $k$ repetitions of the tractography phase, and averaging the numbers of fibers for each edge on the $k$ runs.

 For $i=1,2,\ldots,10$, we have generated independent $k$ instances, and averaged these $k$ fiber numbers for each edge. Next, we have thrown out those edges, which were not present in all the ten copies of the averaged graphs. Now, for each remaining edge $\{u,v\}$ of the graph $G$, we computed the average fiber number values over $k$ repetitions: one average value $w^{(k)}_i(u,v)$ for each $i$ in ${G_k}_i$, for $i=1,2,\ldots,10$. For readability, we omit $(u,v)$ from $w^{(k)}_i(u,v)$ in what follows.

For these ten $w^{(k)}_i$ values we computed the relative standard deviation (also called coefficient of variation) of the ten $w^{(k)}_i$ values:

$$c_v(w^{(k)})={\sigma(w^{(k)})\over\mu(w^{(k)})}, \eqno{(1)} $$  
		where
$$\mu(w^{(k)})={ 1\over10}\sum_{i=1}^{10}w^{(k)}_i, \ \   \sigma(w^{(k)})=\sqrt{{1\over9}\sum_{i=1}^{10} (w^{(k)}_i-\mu(w^{(k)}))^2}
 \eqno{(2)}  $$

Figure 1 displays the change of the relative standard deviation of the fiber number of a given edge (the edge, connecting vertex 17 and vertex 21 in the 463-vertex resolution in the case of subject No. 901038) for $k=1,2,\ldots,50$.

\begin{figure}[H]
	\begin{center}
		\includegraphics[width=8cm]{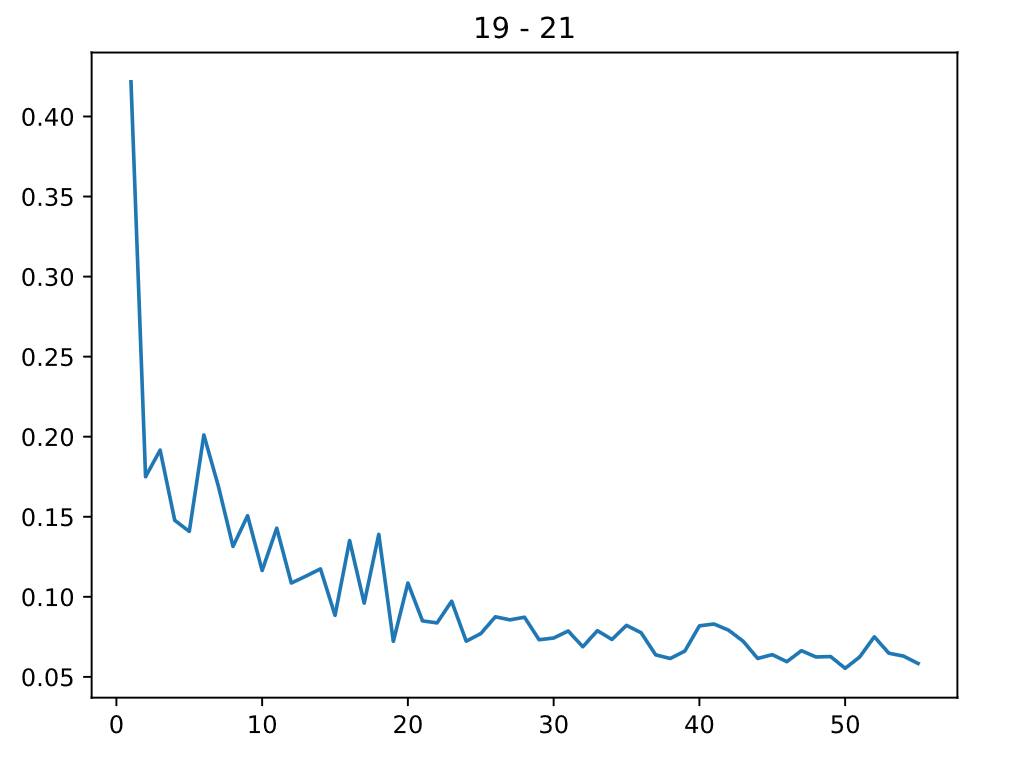}
		\caption{The change of the relative standard deviations of the edge, connecting vertex 17 and vertex 21 in the 463-vertex resolution in the case of subject No. 901038, for $k=1,2,\ldots,50$.}
	\end{center}
\end{figure}

Figure 2 shows the change of the relative standard deviations, averaged for all edges as a function of $k$, in the case of a given braingraph, in 234-vertex resolution. Supporting Figures 1, 2, 3 and 4 show the same in graphs of different resolutions.

 \begin{figure}[H]
 	\begin{center}
 		\includegraphics[width=16cm]{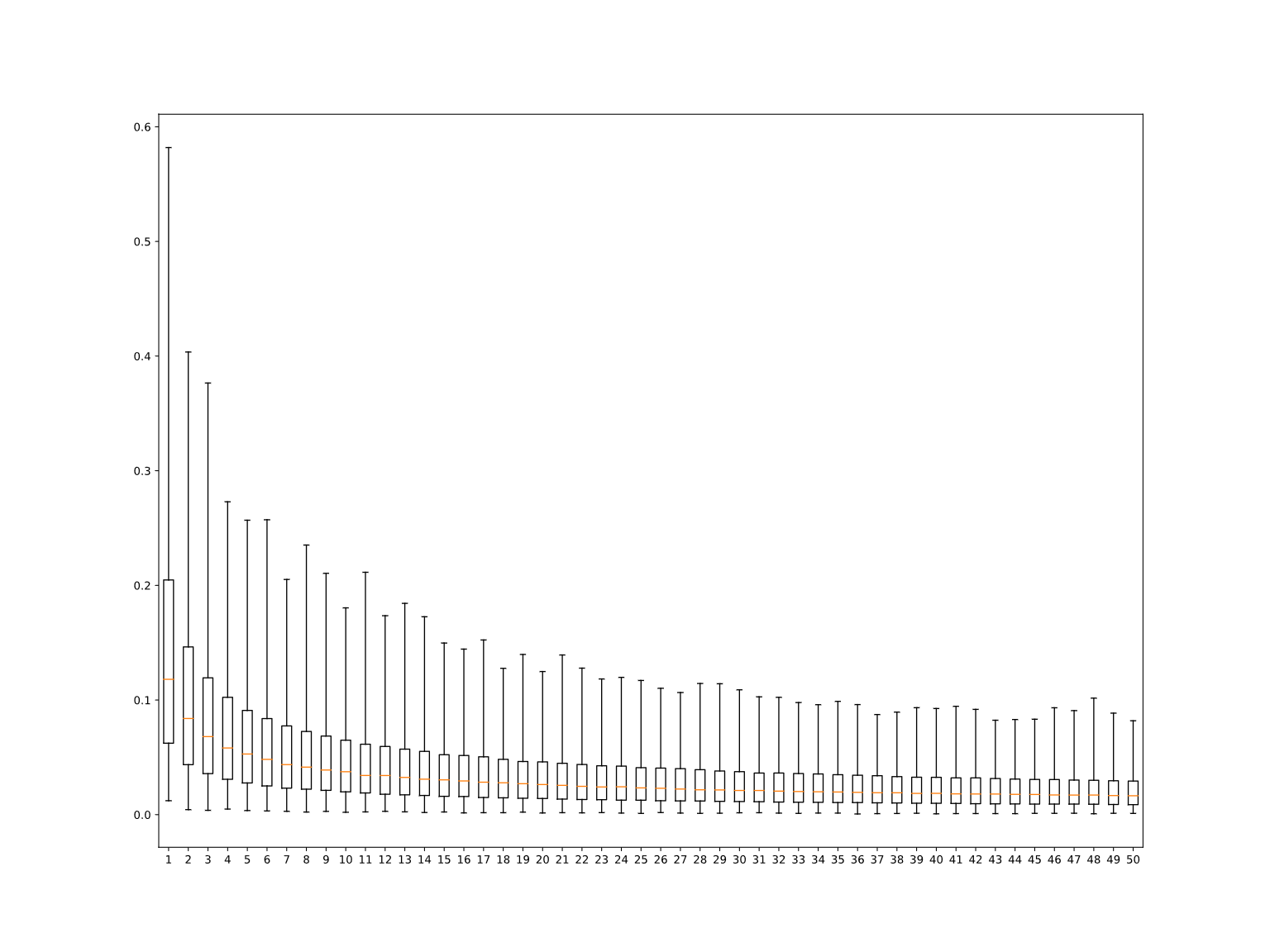}
 		\caption{The change of the relative standard deviations, averaged for all edges as a function of  $k=1,2,\ldots,50$, in the case of the connectome of subject No. 300618, in 234-vertex resolution. The medians of the relative standard deviations are visualized by red horizontal lines, while the boxes show the middle-half of the datapoints: under the box there are the lower quarter-, above the box the upper quarter of the data points. The solid lines show the whole spread of the data points.}
 	\end{center}
 \end{figure}

Based on Figure 2 (and the related figures for other resolutions and subjects, cf. Supporting Figures 1, 2, 3 and 4), we have chosen the $k=10$ value for repetitions as a good trade-off between deviation and practical computability.

\section*{Acknowledgments}
Data were provided in part by the Human Connectome Project, WU-Minn Consortium (Principal Investigators: David Van Essen and Kamil Ugurbil; 1U54MH091657) funded by the 16 NIH Institutes and Centers that support the NIH Blueprint for Neuroscience Research; and by the McDonnell Center for Systems Neuroscience at Washington University. VG and BV were partially supported by the VEKOP-2.3.2-16-2017-00014 program, supported by the European Union and the State of Hungary, co-financed by the European Regional Development Fund, and the NKFI-127909
 grant of the National Research, Development and Innovation Office of Hungary. VG and BV were supported in part by the EFOP-3.6.3-VEKOP-16-2017-00002 grant, supported by the European Union, co-financed by the European Social Fund.
\bigskip 

\noindent Conflict of Interest: The authors declare no conflicts of interest.

\section*{Author Contribution} BV constructed the image processing system, computed the braingraphs, invented the method of optimizing the repeated tractographies, and prepared the figures, VG has secured funding, initiated the study, analyzed data and wrote the paper.



\section*{Supporting Figures}

Supporting Figures 1, 2, 3 and 4 visualize the change of the relative standard deviations, averaged for all edges as a function of  $k=1,2,\ldots,50$, in the case of the connectome of subject No. 300618, in 83, 129, 463 and 1015-vertex resolutions, respectively. The medians of the relative standard deviations are visualized by red horizontal lines, while the boxes show the middle-half of the data points: under the box there are the lower quarter-, above the box the upper quarter of the data points. The solid lines show the whole spread of the data points.

\renewcommand\thefigure{\arabic{figure}}

\setcounter{figure}{0}

\subsection*{Supporting Figure 1}

	\includegraphics[width=16cm]{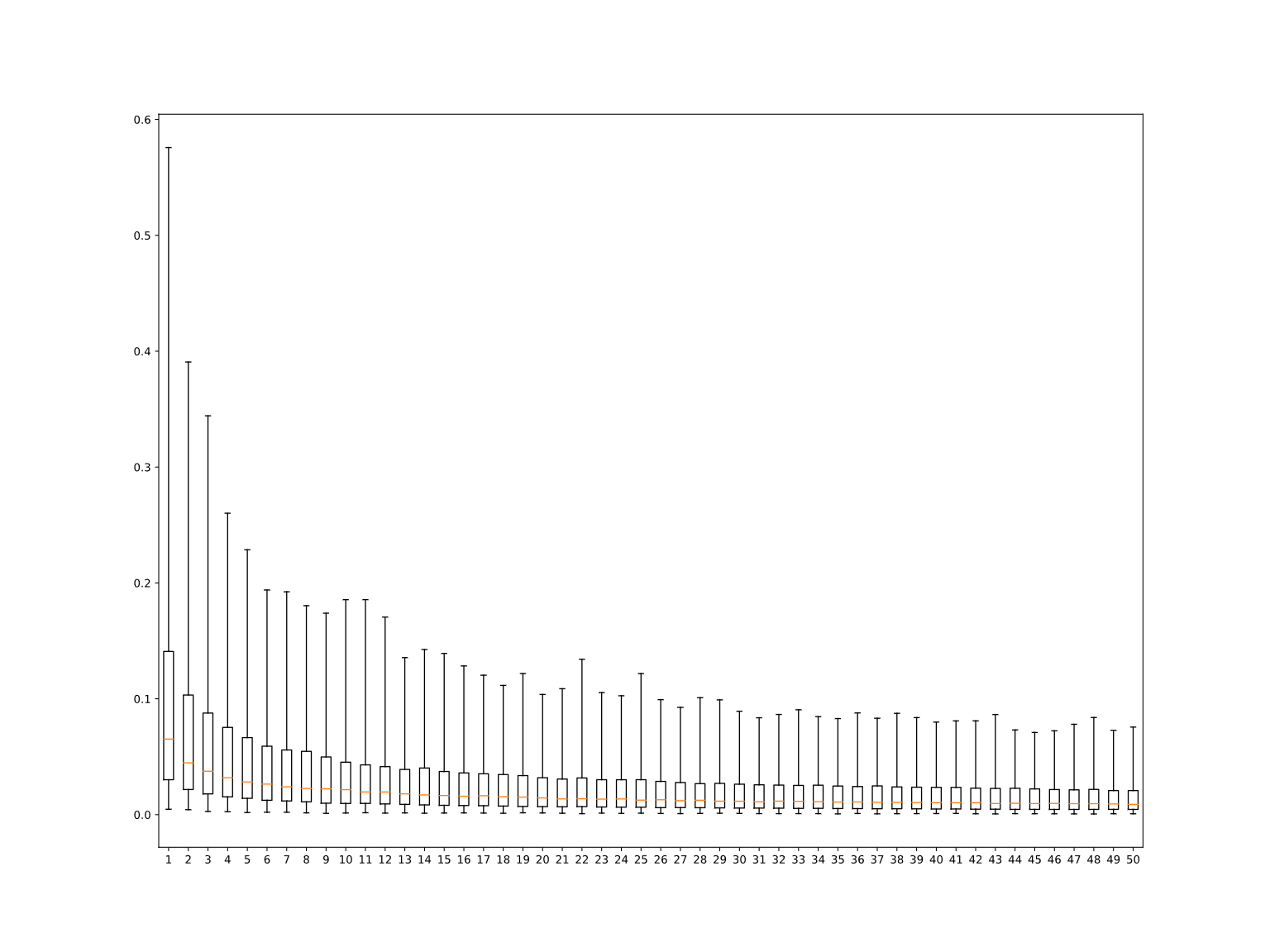}
	
\subsection*{Supporting Figure 2}
	
	\includegraphics[width=16cm]{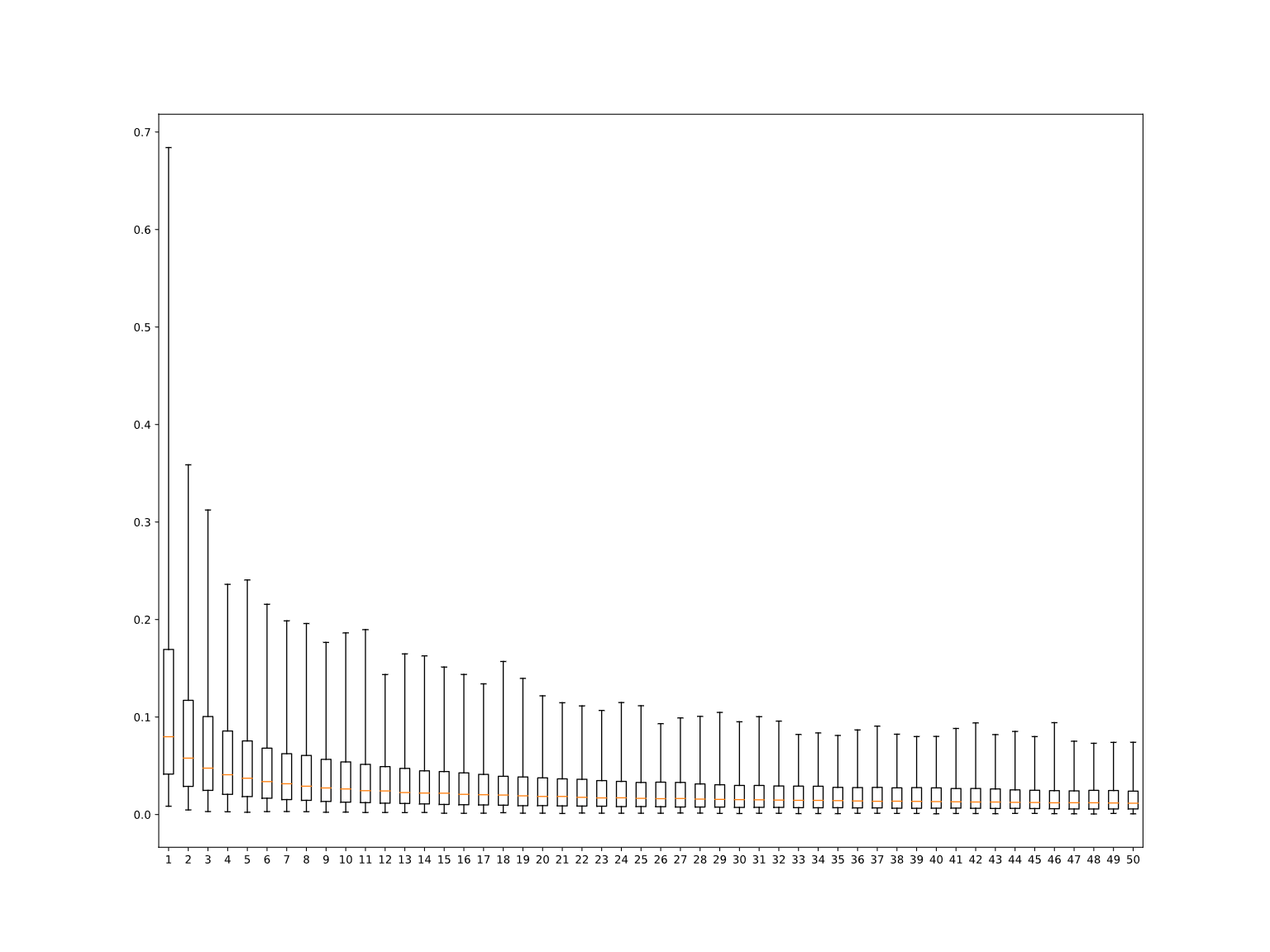}
	
\subsection*{Supporting Figure 3}
	
	\includegraphics[width=16cm]{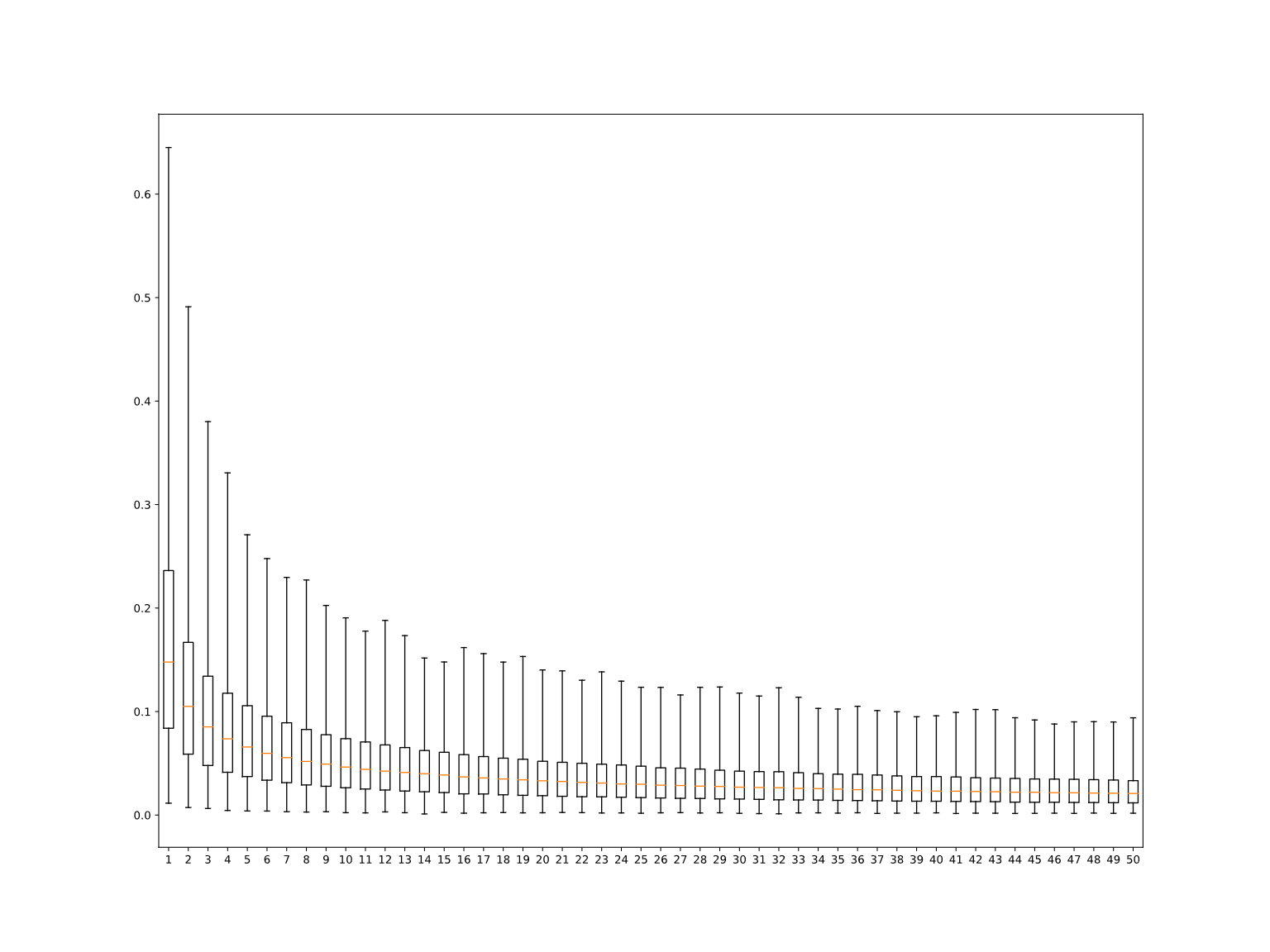}
	
\subsection*{Supporting Figure 4}
	
	\includegraphics[width=16cm]{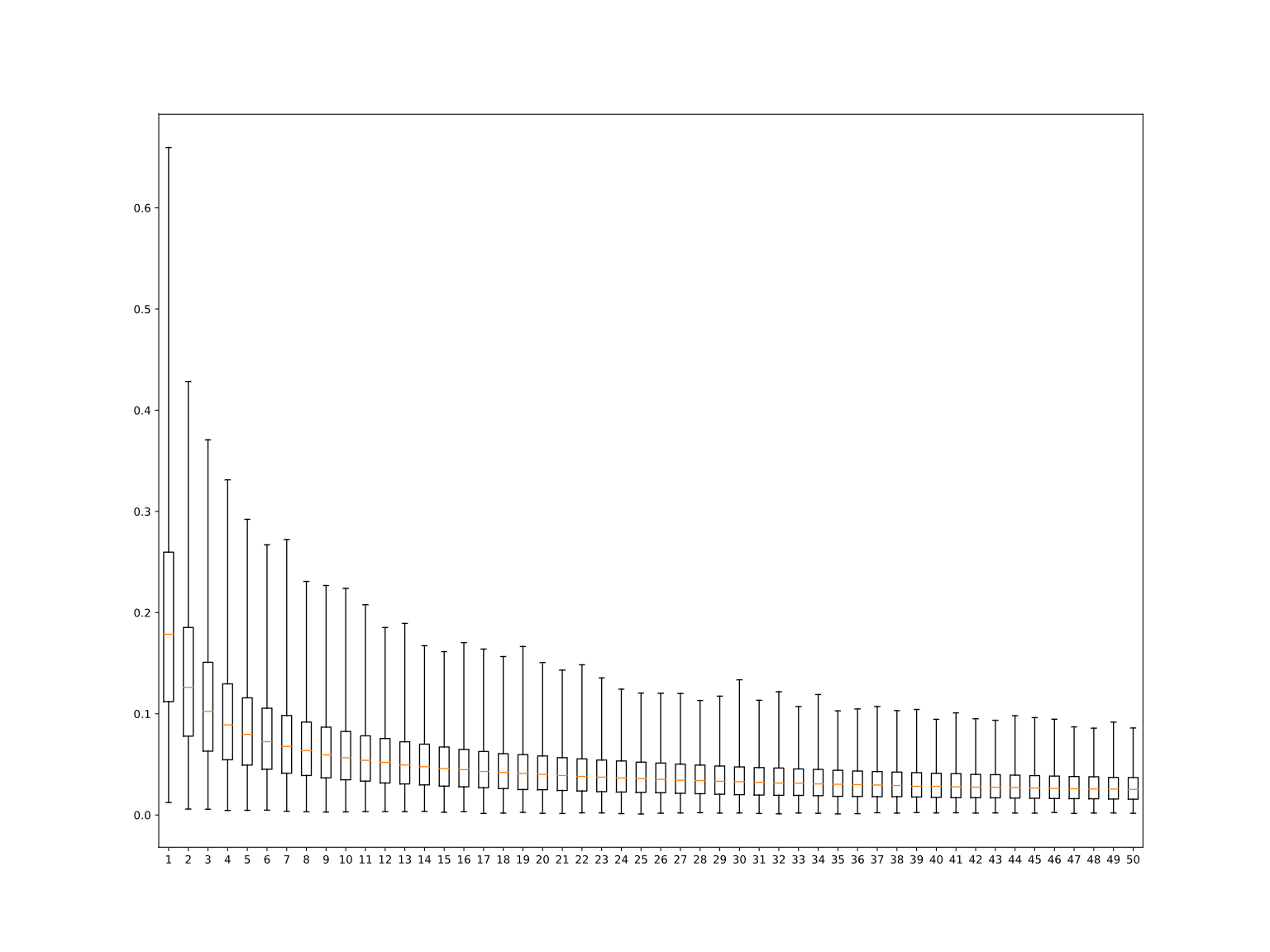}

\end{document}